\documentclass[11pt]{article}

\usepackage{amsmath,amssymb,amsfonts}

\usepackage{times}
\usepackage{bm,mathrsfs,dsfont,color}
\usepackage{natbib}
\usepackage{graphicx}
\usepackage{subfigure}%,caption}
\usepackage[font=small]{caption}

\usepackage{authblk}

\allowdisplaybreaks

\usepackage{multirow}

\newtheorem{lemma}{Lemma}

\newcommand{\E}{\mathds{E}}

\def\tr{{\rm T}}

\def\var{{\rm Var}}

\usepackage[plain,noend]{algorithm2e}

\makeatletter
\renewcommand{\algocf@captiontext}[2]{#1\algocf@typo. \AlCapFnt{}#2} % text of caption
\def\@algocf@capt@plain{top}
\renewcommand{\algocf@makecaption}[2]{%
  \addtolength{\hsize}{\algomargin}%
  \sbox\@tempboxa{\algocf@captiontext{#1}{#2}}%
  \ifdim\wd\@tempboxa >\hsize%     % if caption is longer than a line
    \hskip .5\algomargin%
    \parbox[t]{\hsize}{\algocf@captiontext{#1}{#2}}% then caption is not centered
  \else%
    \global\@minipagefalse%
    \hbox to\hsize{\box\@tempboxa}% else caption is centered
  \fi%
  \addtolength{\hsize}{-\algomargin}%
}
\makeatother

\begin{document}

%% Here are the title, author names and addresses
\makeatletter
\renewcommand\section{\@startsection {section}{1}{\z@}%
                                   {-3.5ex \@plus -1ex \@minus -.2ex}%
                                   {2.3ex \@plus.2ex}%
                                   {\centering\normalfont\large\scshape}}
                                   
\renewcommand\subsection{\@startsection {subsection}{1}{\z@}%
                                   {-3.5ex \@plus -1ex \@minus -.2ex}%
                                   {2.3ex \@plus.2ex}%
                                   {\centering \normalfont \scshape}}                                   
\makeatother

\title{Pairwise likelihood inference \\ for the multivariate ordered probit model}

\author[1]{Martina Bravo}

\author[2]{Antonio Canale}

\affil[1]{Department of Mathematics,  University of Torino}
\affil[2]{Department of Statistical Sciences, University of Padova}

\date{}

\maketitle

\abstract{This paper provides a closed form expression for the pairwise score vector for the multivariate ordered probit model. This result has several implications in likelihood-based inference. It is indeed used both to speed-up gradient based optimization routines for point estimation, and to provide a building block to compute standard errors and confidence intervals by means of the Godambe matrix. }

{\center \textbf{Keywords: }}
composite likelihood; Godambe information matrix; pairwise score function

\maketitle

\section{Introduction}

Multivariate ordinal categorical outcomes \citep{agres2002} arise in many fields of application such as political science, marketing research, educational assessment, and psychometry among others. A common approach consists in considering that the $q$ observed categorical data are related to continuous latent variables. The relation between the latent continuous variables and the observed categorical variables is usually induced by means of thresholds partitioning the latent sample space into a series of regions corresponding to each ordinal category. Popular distributions for the latent variables are the  logistic and the Gaussian distributions, leading to the ordered logit and probit models, respectively.

In this paper, we deal with inferential issues related to the multivariate ordered probit class of model following a likelihood-based approach. Specifically, since the multivariate ordered probit model has clear computational problems related to the calculation of a $q$ dimensional integral for each single likelihood contribution, we exploit an approximate approach based on a simple likelihood belonging to the class of composite likelihoods \citep{lind:1988, varin:etal:2011}: the pairwise likelihood  \citep{cox:reid:2004}. The use of the pairwise likelihood  for the ordered probit model has been shown to have clear advantages in many situations \citep{de2005pairwise, varin2006pairwise, tonyclo, viennesi}.
In particular, in the context of point estimation, \citet{tonyclo} showed a dramatic improvement in the computational time and only moderate bias if compared to a standard likelihood approach. This comparison, however, was only possible for moderate values of $q$ as the standard likelihood approach is not practically feasible in high dimensions. \citet{viennesi} confirm this evidence while showing that a triplewise likelihood does not lead to substantial improvement further endorsing the pairwise approximation. In both of the aforementioned papers, the authors try to go beyond point estimation and provide significance testing on some parameters. There is however a clear difficulty in obtaining inferential results beyond point estimation with the approach at hand: an analytic expression for the covariance matrix of the pairwise likelihood estimator is not available. This is mainly due to the difficulties in calculating the Godambe matrix, an estimator for the asymptotic precision of the maximum pairwise likelihood estimator. Indeed, the latter involves the calculation of the derivative of the pairwise log-likelihood, i.e. the derivative of the logs of bivariate integrals of Gaussian densities with respect to the limits of integration and the latent correlation coefficients. To bypass this problem, possible solutions consist in calculating the the derivative of the pairwise log-likelihood numerically or in avoiding asymptotic arguments and instead using a bootstrap-based approach \citep{tonyclo}. 
Unfortunately both approaches are computationally demanding in moderate and high dimensions. While \citet{viennesi} mention that the estimates standard errors may be obtained  ``either analytically or by numerical differentiation'' (see page 8 of \citet{viennesi}) in their paper they do not provide any analytical result.

The analytical expression of the pairwise score vector would be hence of paramount interest both to speed-up gradient based optimization routines for point estimation, and in order to compute both the sensitivity and variability matrices involved in the Godambe's matrix computation. This paper fills this gap and presents the analytical expression of the pairwise score vector of the multivariate ordered probit model.

The paper is organized as follows: the next section introduces the multivariate ordered probit model.  Section~\ref{martina} ---after a brief review of the pairwise likelihood approach to inference--- reports the main result of this paper, i.e. the closed form expression for the pairwise score vector. In Section~\ref{sec:time} some numerical and computational considerations are reported. Section~\ref{sec:illustration} reports a simulation experiment to assess the performance of the pairwise approach to perform confidence interval estimation by means of an empirical approximation of the Godambe's matrix. Section~\ref{conclusion} concludes the paper discussing some ongoing research.

%##########################################################
\section{The multivariate ordered probit model}
\label{model}
%##########################################################

Let $Y_i=(Y_{i1},\ldots, Y_{iq})^{\tr}$, $Y_{ij}  \in \{1,2,\ldots,K\}$ for $j = 1, \dots, q$ be a $q$-dimensional ordinal categorical random vector with joint distribution 
depending on some unknown parameter $\theta$, with $i =1, \dots, n$ defining a collection of iid random vectors.  The ordered probit model assumes that, for each $i =1, \dots, n$, there exists a latent random vector ${ Z_i}=(Z_{i1},\ldots, Z_{iq})^{\tr} $, with  ${ Z_i} \stackrel{iid}{\sim} N(0, \Sigma)$ where 
\begin{align*}
\Sigma = \left( \begin{array}{ccccc} 1 & \rho_{1,2} & \dots&\dots& \rho_{1,q} \\ 
		    & 1      & \dots &\dots& \rho_{2,q} \\
		    &  & 1 & \rho_{r,s} & \vdots \\
		    &  &  & 1& \rho_{q-1,q} \\
		    &  &  & & 1 \end{array}
\right)
\end{align*}
is an unknown $q$-dimensional positive definite correlation matrix. In observing $y_{ij}$, the following relation is assumed
\begin{align}
	y_{ij} = k  \text{ if and only if } z_{ij} \in (a_{k-1}(j), a_k(j)], \notag
\end{align}
where for every $j=1,\dots,q$, $\{a_k(j)\}_{k=0,\dots,K}$ is a sequence of real numbers such that
$
-\infty = a_0(j) < a_1(j) < \dots < a_{K-1}(j) < a_K(j) = \infty,
$
defining a disjoint partition of $\mathbb{R}$.
The real vector $a(j)=\{a_1(j),\dots,a_{K-1}(j)\}$ is then a $(K-1)$-dimensional vector of threshold parameters referred to the $j$-th marginal. 

Under the ordered probit model assumption, the likelihood contribution of a single observation is then proportional to the joint probability
\begin{align*}
\text{pr}(Y_{i1}=y_{i1}, \dots , Y_{iq}=y_{iq})= 
\int_{a_{y_{i1}-1}(1)}^{a_{y_{i1}}(1)} \dots \int_{a_{y_{iq}-1}(q)}^{a_{y_{iq}}(q)}
\phi_{\Sigma}(z_{1}, \dots, z_{q}) \, dz_{1} \dots  \,dz_{q},
\end{align*}
where $\phi_{\Sigma}(\cdot)$ denotes the multivariate Gaussian distribution with zero mean and variance $\Sigma$. 
Let $\theta$ be the joint vector of parameters, i.e.
$$
\theta=(\rho_{1,2}, \dots, \rho_{q-1,q}, a_1(1),\dots, a_{K-1}(1),
										 a_1(2),\dots, a_{K-1}(2), 
										 a_1(q),\dots, a_{K-1}(q))^T,
$$
then, the log-likelihood function is given by 
\begin{align}
\label{pairfulllik}
\ell(\theta; y) & = \sum_{i=1}^{n}   \text{log}  \left\{ 
\int_{a_{y_{i1}-1}(1)}^{a_{y_{i1}}(1)} \dots \int_{a_{y_{iq}-1}(q)}^{a_{y_{iq}}(q)}
\phi_{\Sigma}(z_{1}, \dots, z_{q}) \, dz_{1} \dots  \,dz_{q}
\right\},
\end{align}
where $y$ denotes all the observed sample. The total number of model parameters is given by the sum of the number of thresholds ($(K-1)\times q$) plus number of latent correlation coefficients ($q(q-1)/ 2$).
As already discussed, equation \eqref{pairfulllik} involves a $q$-dimensional integral for each observation and hence it is not easily manageable for moderate and high values of $q$.

\section{Pairwise inference for the multivariate probit model}
\label{martina}

\subsection{Pairwise likelihood inference}

The pairwise likelihood is a likelihood constructed from bivariate marginals.  For our multivariate ordered categorical data, the pairwise log-likelihood is
\begin{align}
\ell^P(\theta;y)= &\sum_{i=1}^{n}\sum_{r=1}^{q-1}\sum_{s=r+1}^{q}  \text{log} 
\bigl\{ \mbox{pr}(Y_{ir}=y_{ir} , Y_{is}=y_{is})    \bigr\} \notag	\\
= & \sum_{i=1}^{n}\sum_{r=1}^{q-1}\sum_{s=r+1}^{q}  \text{log} 
\left\{ \int_{a_{y_{ir}-1}(r)}^{a_{y_{ir}}(r)} \int_{a_{y_{is}-1}(s)}^{a_{y_{is}}(s)}   
\phi_{\Sigma(\rho_{r,s})}(z_{r},z_{s}) \, dz_{r}\,dz_{s}    \right\}, 
\label{pairloglik}
\end{align}
where   $\Sigma(\rho)$ denotes the $2 \times 2 $ correlation matrix with off-diagonal entries equal to $\rho$.
As already discussed, the pairwise approach is particularly appealing because it substitutes the computational challenges related to a $q$ dimensional numerical integration with simpler bivariate integrals.  %Similar approaches has been considered by \citet{de2005pairwise, varin2006pairwise, tonyclo}. FORSE SPOSTARE QUESTO SOPRA.

Generally, composite likelihood inferential procedures shares many properties of standard likelihood methods. For example, the pairwise score vector, $u^P(\theta;y)=\partial\ell^P(\theta;y)/\partial\theta,$ is still unbiased being
the sum of score vectors based on the likelihood contribution of each pair of observations.
Under regularity conditions, the maximum pairwise likelihood estimator $\hat{\theta}$ ---obtained either by maximizing  $\ell^P(\theta;y)$ numerically or by solving the pairwise likelihood equations $u^P(\theta;y)=0$--- is consistent and has asymptotic Gaussian distribution
\begin{align*}
\sqrt{n}(\hat\theta-\theta) \dot\sim N(0, G(\theta)^{-1}).
\end{align*}
In equation above $G(\theta)=H(\theta)J(\theta)^{-1}H(\theta)$ is the Godambe information matrix where
$$
H(\theta)=\E\{ -\partial u^P(\theta;y)/\partial\theta\} \mbox{ and } 
J(\theta)=\var \{u^P(\theta;y)\}
$$
 are the sensitivity and variability matrices, respectively.
The sample estimate of the variability matrix is simply
\begin{align}
\hat{J}(\theta)=\frac{1}{n} \sum_{i=1}^n u^P(\theta;y_i)  u^P(\theta;y_i)^T \Big|_{\theta=\hat{\theta}}, 
\label{Jhat}
\end{align}
while the sample estimate of the sensitivity matrix is given by 
\begin{align*}
\hat{H}(\theta)=-\frac{1}{n} \sum_{i=1}^n \frac{\partial u^P(\theta;y_i)}{\partial\theta}\Big|_{\theta=\hat{\theta}}.
\end{align*}
However, the calculation of the matrix of the derivatives of the score can be avoided by exploiting the second Bartlett identity, which is still valid as the pairwise likelihood is made of proper likelihood contributions. This yields the alternative expression 
\begin{align}
\hat{H}(\theta)=\frac{1}{n} \sum_{i=1}^{n}\sum_{r=1}^{q-1}\sum_{s=r+1}^{q}  u(\theta;y_{ir},y_{is})  u(\theta;y_{ir},y_{is})^T \Big|_{\theta=\hat{\theta}},
\label{Hhat}
\end{align}
where $u(\theta;y_{ir},y_{is})$ is a  suitable adaptation of the score vector of a bivariate Gaussian likelihood with data $(y_{ir},y_{is})^T$. This adaptation is needed to match the dimension of the full vector of parameters $\theta$, i.e. $q(q-1)/2 + (K-1)q$ as each bivariate Gaussian likelihood involves $1+2(K-1)$ parameters only. Specifically, if a component of $\theta$ is not present in the specific  bivariate Gaussian likelihood, a zero is assigned to the related entry of $u(\theta;y_{ir},y_{is})$.

\subsection{The pairwise score vector}

In this section we are going to obtain closed-form expressions for each element of the pairwise score vector for the pseudolikelihood \eqref{pairloglik}. To our knowledge, this is the first time that such a result is analytically available. In addition, this result is of paramount interest both to speed-up gradient based optimization routines and, even more, to provide sample estimates for the sensitivity and variability matrices used to obtain the asymptotic variance of the pairwise likelihood estimator. The derivation of the following expressions are carefully described in the Master thesis of \citet{bravo} defended at the University of Turin, Italy.
   
Using linearity in differentiation and the property of the derivation of a composite function, the crucial point related to the calculation of the pairwise score vector is the computation of the derivative of the argument of the logarithm in \eqref{pairloglik}.
To this end, consider the standard bivariate Gaussian cumulative distribution function (CDF)
\begin{align*}
\Phi_2(a,b,\rho)= \int_{-\infty}^{a}\int_{-\infty}^{b} \phi_{\Sigma(\rho)}(u,v) \,du  \,dv,
\end{align*}

and note that the general integral of a bivariate Gaussian density  can be decomposed in the sum of four CDF, namely
\begin{align}
\label{scompos}
\int_{b}^{a}\int_{d}^{c} \phi_{\Sigma(\rho)}(u,v) \,du \,dv= 
\Phi_2(a,c;\rho)-\Phi_2(b,c;\rho)-\Phi_2(a,d;\rho)+
\Phi_2(b,d;\rho).
\end{align}
Hence the arguments of all the logarithms in \eqref{pairloglik} can be easily rewritten according to equation \eqref{scompos} thus allowing to rewrite the pairwise log-likelihood function as
\begin{align}
\label{loglik}
\ell^P(\theta;y) & =
 \sum_{i=1}^{n}\sum_{r=1}^{q-1}\sum_{s=r+1}^{q}  \text{log}  
\biggl\{ \Phi_2\Bigr(a_{y_{ir}}(r),a_{y_{is}}(s),\rho_{r,s}\Bigl) -
\Phi_2\Bigr(a_{y_{ir}-1}(r),a_{y_{is}}(s),\rho_{r,s}\Bigl)  +
\\
& - \Phi_2\Bigr(a_{y_{ir}}(r),a_{y_{is}-1}(s),\rho_{r,s}\Bigl)  + 
\Phi_2\Bigr(a_{y_{ir}-1}(r),a_{y_{is}-1}(s),\rho_{r,s}\Bigl)
\biggr\}.
\end{align}

It is clear that the partial derivatives of the score vector, follow from the partial derivatives of bivariate Gaussian CDFs. To obtain the latter,  consider the following results \citep{Drum}.
\begin{lemma}
\label{lemma}
	The partial derivatives of the bivariate standard Normal CDF with respect to the limits of integration and the correlation coefficients are 
	\begin{align} 
	\label{first}
\frac{\partial \Phi_2(x_1,x_2,\rho)}{\partial x_1}&= \phi(x_1)  \Phi \biggl( \frac{x_2 - \rho x_1}{\sqrt{
			1-\rho^2}} \biggr),\\
	\label{second}
\frac{\partial \Phi_2(x_1,x_2,\rho)}{\partial x_2}&= \phi(x_2)  \Phi \biggl( \frac{x_1 - \rho x_2}{\sqrt{
			1-\rho^2}} \biggr),\\
	\label{third}
	\frac{\partial \Phi_2(x_1,x_2,\rho)}{\partial \rho} &= \phi_{\Sigma(\rho)}(x_1,x_2).
	\end{align}
\end{lemma}

The application of Lemma~\ref{lemma} allows us to easily compute the partial derivatives with respect to all correlation coefficients, i.e. for every $r=1,\dots, q-1$ and $s=r+1,\dots, q$, we have
\begin{align}
\label{partialrho}
\frac{\partial \ell^P(\theta)}{\partial \rho_{r,s}} & =
\sum_{i=1}^{n}  \left\{ \int_{a_{y_{ir}-1}(r)}^{a_{y_{ir}}(r)} \int_{a_{y_{is}-1}(s)}^{a_{y_{is}}(s)}   
\phi_{\Sigma(\rho_{r,s})}(z_{r},z_{s}) \ dz_{r}dz_{s}    \right\}^{-1} \times \notag
\\
&\quad  \times \biggr[\phi_{\Sigma(\rho_{r,s})} \Bigr(a_{y_{ir}}(r),a_{y_{is}}(s)\Bigl)- \ 
\phi_{\Sigma(\rho_{r,s})}\Bigr(a_{y_{ir}-1}(r),a_{y_{is}}(s)\Bigl) \ + \notag
\\
&\quad  -  \phi_{\Sigma(\rho_{r,s})}\Bigr(a_{y_{ir}}(r),a_{y_{is}-1}(s)\Bigl) \ + \ 
\phi_{\Sigma(\rho_{r,s})}\Bigr(a_{y_{ir}-1}(r),a_{y_{is}-1}(s)\Bigl)   \biggl].
\end{align} 
The latter equation contributes to determine the first entries of the score vector $u^P(\theta;y)=\partial \ell^P(\theta;y)/\partial \theta$.

In computing the partial derivatives with respect to the thresholds parameters, the application of Lemma~\ref{lemma} is not trivial. First of all, note that  in \eqref{pairloglik} the extremes of integration still depend on the observations $y_{ij}$. Hence, it is useful to rewrite \eqref{pairloglik}, grouping the pairs of observations sharing the same values $y_{ij}$. To this end, we can first rewrite the pairwise likelihood \eqref{pairfulllik} as
\begin{equation*}
\label{multiv}
\mathcal{L}^P(\theta)=\prod_{l=1}^{K} \ \prod_{m=1}^{K} \ \prod_{r=1}^{q-1}\prod_{s=r+1}^{q}
\mbox{pr}\Bigr(Y_r=l,Y_s=m\Bigl)^{n_{\{r,l\}\{s,m\} }},
\end{equation*}
where $n_{\{r,l\}\{s,m\} }$ is the number of observations with $Y_{ir}=l$ and $Y_{is}=m$, i.e.
$$
n_{\{r,l\}\{s,m\} }=\sum_{i=1}^{n} {1}_{ \{Y_{ir}= \ l, \ Y_{is}= \ m  \}  }.
$$
Clearly the $n_{\{r,l\}\{s,m\} }$ multiplicities are such that
\[
\sum_{l=1}^{K} \sum_{m=1}^{K} n_{\{r,l\}\{s,m\} }=\frac{q(q-1)}{2}n.
\]
The full log-likelihood in \eqref{pairloglik} can be rewritten as 
\begin{equation*}
\begin{split}
\ell^P(\theta) & =  \sum_{l=1}^{K}\sum_{m=1}^{K}\sum_{r=1}^{q-1}\sum_{s=r+1}^{q} n_{\{r,l\}\{s,m\} }  \log
\Bigl\{ \mbox{pr}\Bigr(Y_r=l,Y_s=m\Bigl)    \Bigr\}
\\
& =\sum_{l=1}^{K}\sum_{m=1}^{K}\sum_{r=1}^{q-1}\sum_{s=r+1}^{q} n_{\{r,l\}\{s,m\} }\ \log
\biggl\{ \int_{a_{l-1}(r)}^{a_{l}(r)} \int_{a_{m-1}(s)}^{a_{m}(s)}   
\phi_{\Sigma(\rho_{r,s})}(z_{r},z_{s}) \ dz_{r}dz_{s}    \biggr\}\\
%\end{split}
%\end{equation*}
%which can be written, applying scomposition \eqref{scompos} as
%\\
%\begin{equation*}
%\begin{split}
%\ell^P(\theta)
 & = \sum_{l=1}^{K}\sum_{m=1}^{K}\sum_{r=1}^{q-1}\sum_{s=r+1}^{q} n_{\{r,l\}\{s,m\} } \ \text{log} \ 
\biggl\{ 
\Phi_2\Bigr(a_{l}(r),a_{m}(s),\rho_{r,s}\Bigl) \ - \
\Phi_2\Bigr(a_{l-1}(r),a_{m}(s),\rho_{r,s}\Bigl) \ + \\
&\quad - \ \Phi_2\Bigr(a_{l}(r),a_{m-1}(s),\rho_{r,s}\Bigl) \ +  \ \Phi_2\Bigr(a_{l-1}(r),a_{m-1}(s),\rho_{r,s}\Bigl)  
\biggr\}.
\end{split}
\end{equation*}

We now calculate the partial derivative of the pairwise log-likelihood with respect to $a_k(j)$. This parameter is referred to the $k$-th response level---and hence it is involved when $l=k,k+1$ and $m=k,k+1$---and to the $j$-th marginal dimension. To emphasise the contribution of $a_k(j)$ to the pairwise log-likelihood above, we can rewrite $\ell^P(\theta;y)$ as
\begin{equation}
\label{scomp}
\ell^P(\theta;y)= A_k(j)+B_k(j)+R_k(j),
\end{equation}
where
\\
\begin{equation*}
\begin{split}
A_k(j)  = & \sum_{l=1}^{K}\sum_{m=k}^{k+1} \ \sum_{r=1}^{j-1}
n_{\{r,l\}\{j,m\} } \ \text{log} \ 
\biggl\{ 
\Phi_2\Bigr(a_{l}(r),a_{m}(j),\rho_{r,j}\Bigl) -
\Phi_2\Bigr(a_{l-1}(r),a_{m}(j),\rho_{r,j}\Bigl) +\\
& - \ \Phi_2\Bigr(a_{l}(r),a_{m-1}(j),\rho_{r,j}\Bigl) \ + 
\Phi_2\Bigr(a_{l-1}(r),a_{m-1}(j),\rho_{r,j}\Bigl)  
\biggr\},\\
B_k(j)  = &\sum_{l=k}^{k+1}\sum_{m=1}^{K}\sum_{s=j+1}^{q}
n_{\{j,l\}\{s,m\} } \ \text{log} \ 
\biggl\{ 
\Phi_2\Bigr(a_{l}(j),a_{m}(s),\rho_{j,s}\Bigl) - 
\Phi_2\Bigr(a_{l-1}(j),a_{m}(s),\rho_{j,s}\Bigl) +\\
&\quad -  \Phi_2\Bigr(a_{l}(j),a_{m-1}(s),\rho_{j,s}\Bigl)  + 
\Phi_2\Bigr(a_{l-1}(j),a_{m-1}(s),\rho_{j,s}\Bigl)  
\biggr\},
\end{split}
\end{equation*}
and $R_k(j)$ is a residual part that does not depend on $a_k(j)$.
Note that $A_k(j)$ is the likelihood contribution where the dimension $j$ is fixed as second element in the pairs with levels $k$ and $k+1$, while the first element is free to vary between all the possible previous dimensions and combinations of levels. Similarly, $B_k(j)$ is the likelihood contribution where the dimension $j$ is fixed as first element in the pairs with levels $k$ and $k+1$ and the second is free to vary between all the possible next combinations of levels.

The partial derivative of the pairwise log-likelihood with respect to $a_k(j)$ is hence,
\begin{equation}
\frac{\partial \ell^P(\theta)}{\partial a_k(j)}  = \frac{\partial A_k(j)}{\partial a_k(j)}+
\frac{\partial B_k(j)}{\partial a_k(j)}
\label{partialakj}
\end{equation}
where
\begin{equation*}
\begin{split}
\frac{\partial A_k(j)}{\partial a_k(j)} & = \phi\Bigr(a_k(j)\Bigl) 
\sum_{l=1}^{K}\sum_{r=1}^{j-1} 
\left\{  \Phi \Biggr( \frac{a_l(r)-\rho_{r,j}  a_k(j)}{\sqrt{1-\rho_{r,j}^2}}  \Biggl) -  \Phi \Biggr( \frac{a_{l-1}(r)-\rho_{r,j} a_k(j)}{\sqrt{1-\rho_{r,j}^2}}  \Biggl)
\right\} \times 
\\
&\quad \times \Biggl[ \ n_{\{r,l\}\{j,k\} }  \biggl\{ 
\int_{a_{l-1}(r)}^{a_{l}(r)} \int_{a_{k-1}(j)}^{a_{k}(j)}   
\phi_{\Sigma(\rho_{r,j})}(z_{r},z_{j}) \ dz_{r}dz_{j}  
\biggr\}^{-1} +  \\ 
& \quad\quad\quad - \ n_{\{r,l\}\{j,k+1\} } \biggl\{ 
\int_{a_{l-1}(r)}^{a_{l}(r)} \int_{a_{k}(j)}^{a_{k+1}(j)}   
\phi_{\Sigma(\rho_{r,j})}(z_{r},z_{j}) \ dz_{r}dz_{j}  
\biggr\}^{-1}
\Biggl],
\end{split}
\end{equation*}
and 
\begin{equation*}
\begin{split}
\frac{\partial B_k(j)}{\partial a_k(j)} & = \phi\Bigr(a_k(j)\Bigl) 
\sum_{m=1}^{K}\sum_{s=j+1}^{q} 
\Biggl[  \Phi \Biggr( \frac{a_m(s)-\rho_{j,s} a_k(j)}{\sqrt{1-\rho_{j,s}^2}}  \Biggl) -  \Phi \Biggr( \frac{a_{m-1}(s)-\rho_{j,s}  a_k(j)}{\sqrt{1-\rho_{j,s}^2}}  \Biggl)
\Biggr] 
\\
&\quad \times \Biggl[ n_{\{j,k\}\{s,m\} } \ \biggl\{ 
\int_{a_{k-1}(j)}^{a_{k}(j)} \int_{a_{m-1}(s)}^{a_{m}(s)}   
\phi_{\Sigma(\rho_{j,s})}(z_{j},z_{s}) \ dz_{j}dz_{s}  
\biggr\}^{-1} \  +  \\ 
& \quad\quad\quad- \ n_{\{j,k+1\}\{s,m\} }\biggl\{ 
\int_{a_{k}(j)}^{a_{k+1}(j)} \int_{a_{m-1}(s)}^{a_{m}(s)}   
\phi_{\Sigma(\rho_{j,s})}(z_{j},z_{s}) \ dz_{j}dz_{s}  
\biggr\}^{-1}
\Biggl].
\end{split}
\end{equation*}
Note that for $j=1$ and $j=q$, equations above  simplify to 
\[
\frac{\partial \ell^P(\theta)}{\partial a_k(1)} =\frac{\partial B_k(1)}{\partial a_k(1)}, \quad \quad 
\frac{\partial \ell^P(\theta)}{\partial a_k(q)} =\frac{\partial A_k(q)}{\partial a_k(q)}.
\]

\subsection{Numerical and computational considerations}
\label{sec:time}

\begin{figure}
	\centering
	\subfigure[]{\includegraphics[width=0.44\textwidth]{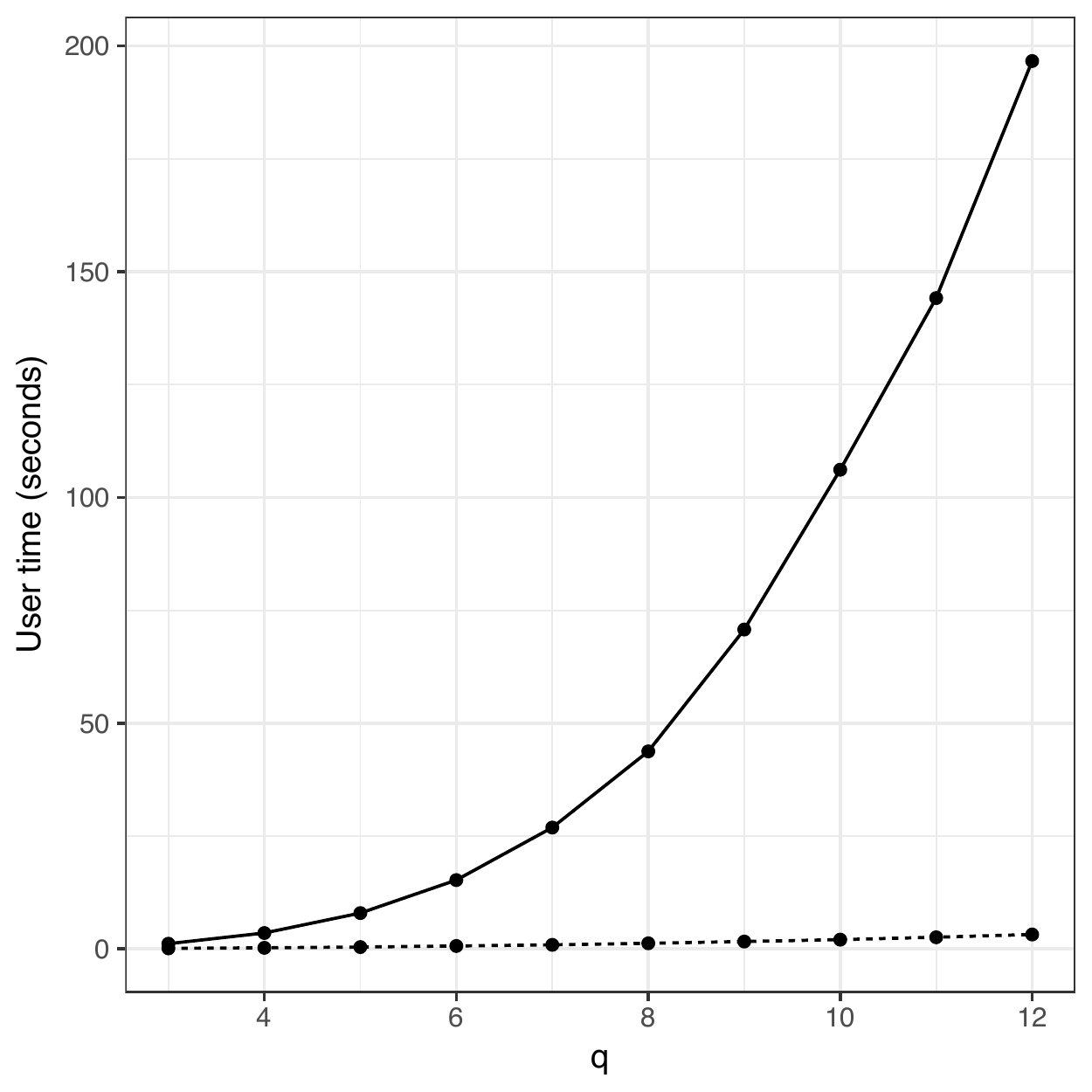}}
	\subfigure[]{\includegraphics[width=0.44\textwidth]{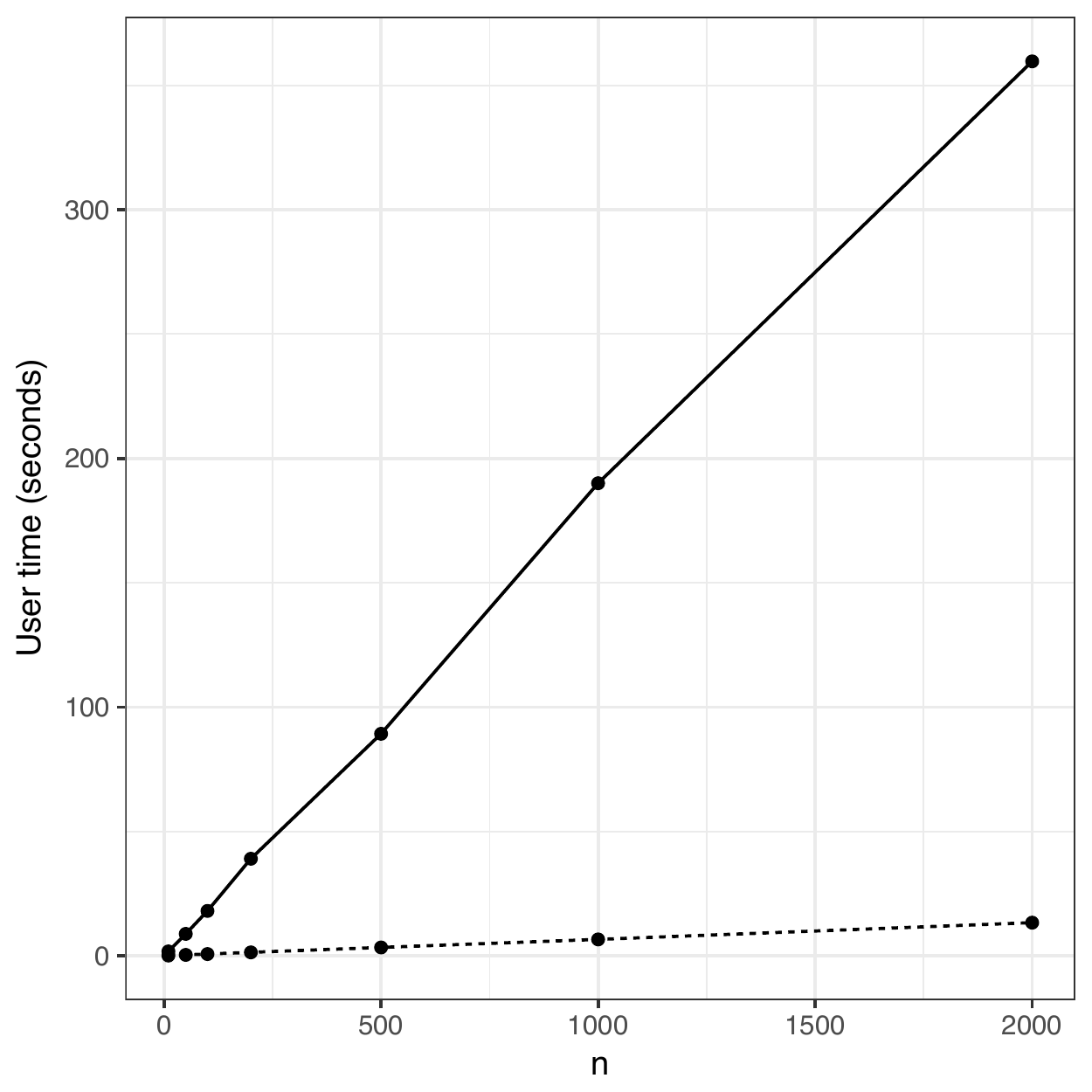}}
	\caption{Execution time of the numerical gradient using the \texttt{grad} function in the R library \texttt{numDeriv} (---) and of the function using the analytical score vector (-  -  -) for (a) different dimensions $q$ (sample size $n=50$ and $K=5$) and (b) different sample size $n$ ($q=5$ and $K=5$)}\label{fig:1}
\end{figure}

The analytical expression of the pairwise score vector equals its numerical equivalent---computed through the \texttt{grad} function in the R library \texttt{numDeriv}. However, its computational burden is dramatically lower. Figure~\ref{fig:1} compares the execution times of a direct calculation of the gradient of the pairwise log-likelihood with the analytical expression of the score for different values of $q$ and fixed sample size and number of levels ($n=50$ and $K=5$). The improvement of our result is dramatic. For instance when $q=12$, the user time to numerically evaluate the score is more that 60 times higher than its analytic counterpart. Similar results are obtained comparing the approaches for growing values of $n$ ---and fixed $q=5$ and $K=5$. The computational costs behave linearly in  both cases, but at a slower rate for the routine using the analytical expression.

As discussed in \cite{tonyclo} and \cite{varin2006pairwise}, the optimization of the pairwise likelihood function, to compute point estimates of the parameters, is performed through the quasi-Newton BFGS algorithm. To assure the ordering of the thresholds $a_0(j) < \dots < a_K(j)$, within the optimization routine, it is useful to define for  $j=1,\dots, q$ and $k=2,\dots, K-1$, the new parameters
\begin{equation}
\label{reparamet}
\delta_k(j) = \log(a_k(j)-a_{k-1}(j)),
\end{equation}
and express the pairwise log-likelihood as a  function of these new parameters. In order to speed up the optimization procedure, we can provide the analytical expression of the score for the model under the new parameter
\begin{equation}
\psi  = (\rho_{1,2}, \dots, \rho_{q-1,q}, \delta_1(1),\dots,\delta_{K-1}(1), \dots ,\delta_1(q), \dots,
\delta_{K-1}(q))^T
\end{equation}
identified by the bijective map $\psi = g(\theta)$. Here $g(\cdot)$ is a vector-to-vector function mapping each correlation coefficient and each $a_1(j)$ with the identity function, and all the remaining threshold parameters with \eqref{reparamet}. The pairwise score vector for the new parameter is simply 
\begin{equation}
\label{form}
u^P(\psi)^T= \frac{\partial \ell^P(\psi)}{\partial \psi} = 
[u^P(g^{-1}(\psi))]^T %\frac{\partial \ell^P(\psi)}{\partial g^{-1}(\psi)}
\cdot  \frac{\partial g^{-1}(\psi)}{\partial \psi},
\end{equation}
where the matrix of partial derivatives of the inverse of $g(\cdot)$ with respect to $\psi$ is the following block matrix 
\begin{equation*}
\frac{\partial g^{-1}(\psi)}{\partial \psi}=
\begin{pmatrix}
 \mbox{I}_{q(q-1)/2} & &&& \\
 &\Delta_1   \\
 & & \Delta_2\\
&&& \ddots \\ 
 & &&& \Delta_q\\\end{pmatrix},
\end{equation*}
where $\mbox{I}_{q(q-1)/2}$ is the identity matrix of dimension ${q(q-1)/2}$, and each $\Delta_j$ is the submatrix of partial derivatives of $g^{-1}(\cdot)$ with respect to the new parameters $\delta_1(j), \dots, \delta_{K-1}(j)$, i.e.
\begin{equation*}
\Delta_j = 
\begin{pmatrix}
 1 & 0 & 0 & \dots & 0 & 0  \\
 1 & e^{\delta_2(j)} & 0 & \dots & 0 & 0 \\
  1 & e^{\delta_2(j)} & e^{\delta_3(j)} & \dots & 0 & 0 \\
 \vdots & \vdots & \vdots & \ddots & \vdots & \vdots \\ 
  1 & e^{\delta_2(j)} & e^{\delta_3(j)} & \dots & e^{\delta_{K-2}(j)} & 0 \\
  1 & e^{\delta_2(j)} & e^{\delta_3(j)} & \dots & e^{\delta_{K-2}(j)} & e^{\delta_{K-1}(j)} \\
\end{pmatrix}.
\end{equation*}
Outside the block diagonal, each entry of the matrix of partial derivatives of the inverse of $g(\cdot)$ with respect to $\psi$ is zero.

\begin{figure}
	\centering
	\subfigure[]{\includegraphics[width=0.44\textwidth]{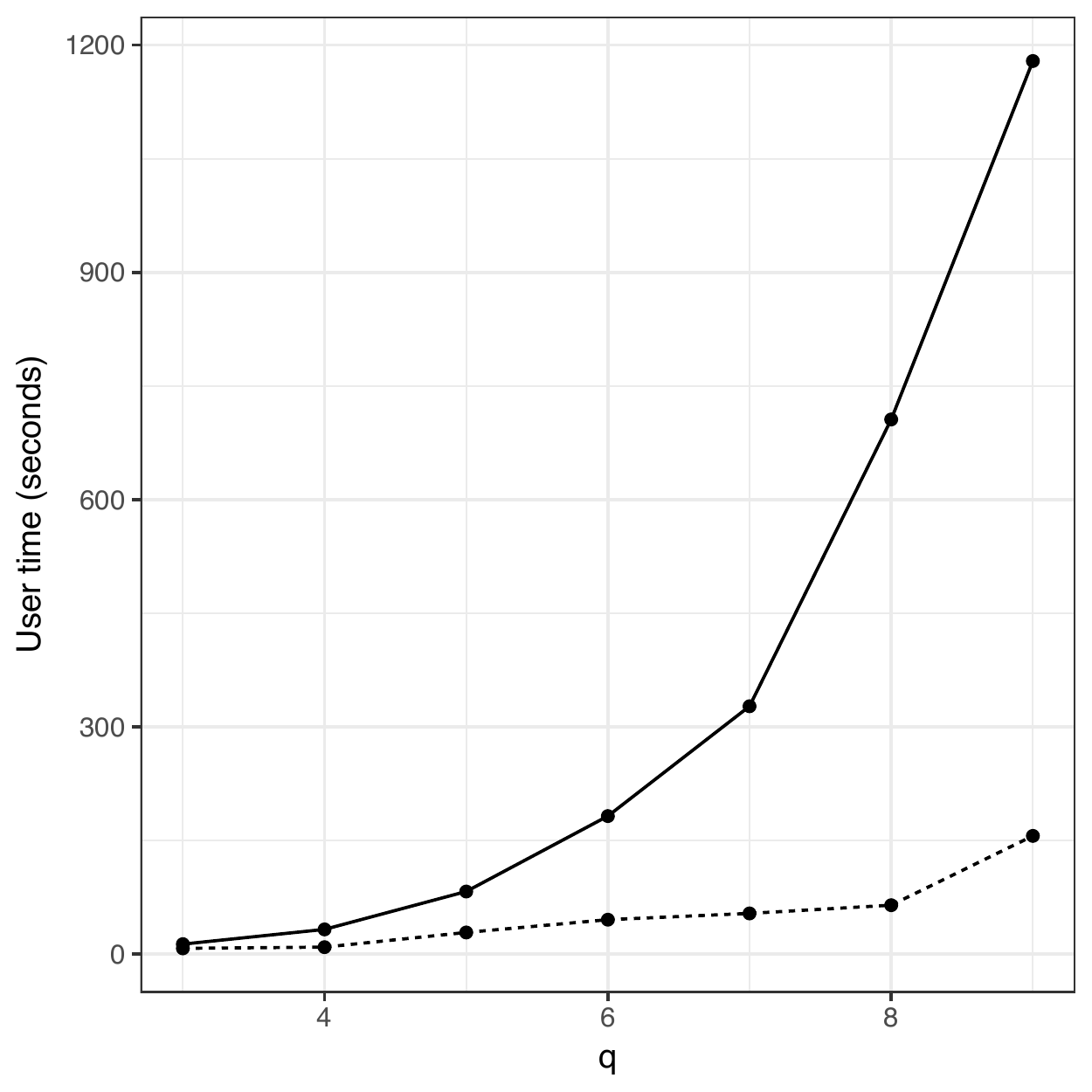}}
	\subfigure[]{\includegraphics[width=0.44\textwidth]{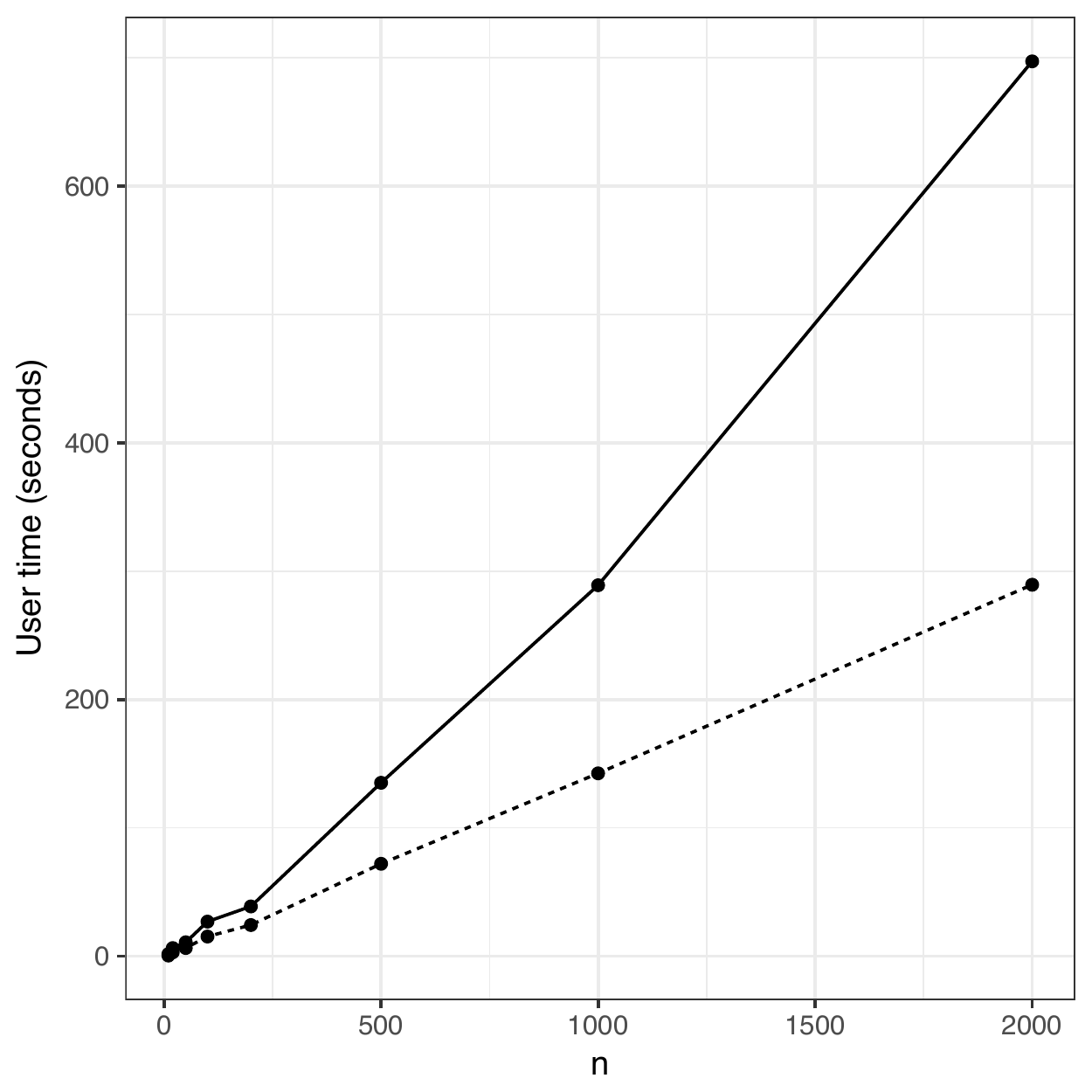}}
	\caption{Execution time of the \texttt{nlminb} procedure without gradient (---) and using the analytical score vector (-  -  -) for 
	(a) different dimensions $q$ (sample size $n=50$ and $K=5$) and (b) different sample size $n$ ($q=5$ and $K=5$).}\label{fig:2}
\end{figure}

Figure \ref{fig:2} compares the execution times of the numerical optimization of the pairwise log-likelihood ---performed with the \texttt{nlminb} function of R--- providing the analytical expression of the score with the numerical optimization in which the gradient is obtained numerically. Data are simulated with different values of dimensions $q$, with $n=50$ and $K=5$ (panel a) and with different sample sizes $n$, fixing $q=3$ and $K=5$ (panel b). In both situations the improvement is considerable. For instance with $q=9$, the optimization with the analytical expression of the gradient performs in less than 3 minutes versus about 20 minutes for the standard procedure.
The use of the algebrical expression obtained in this section dramatically lowers the single pseudolikelihood optimization.

\section{Simulation study}
\label{sec:illustration}

In this section we describe a simulation study conducted in order to assess the performance of the proposed approach in terms of point and confidence interval estimation. 
For confidence interval estimation, we consider the Wald-type confidence interval for all the $q(q-1)/2$ polychoric correlations coefficients. These are defined by the subset of $[-1,1]$ with extremes
\[
\hat{\rho}_{r,s} \pm z_{1-\alpha/2} \ \mbox{se}(\hat{\rho}_{r,s}),
\]
where $z_{1-\alpha/2}$ is the quantile of a standard Gaussian distribution of level $1-\alpha/2$, $\hat{\rho}_{r,s}$ is the pairwise maximum likelihood estimates for ${\rho}_{r,s}$ and $\mbox{se}(\hat{\rho}_{r,s})$ it the appropriate element of the diagonal of the inverse of the Godambe's matrix. 

In fact, the pairwise approach for the ordered probit model has already proved to provide satisfying results in terms of bias and variance for point estimation \cite{tonyclo} \cite{viennesi} but reporting details on point estimation provides additional insights on the performance in terms of confidence interval estimation.

We compute point and interval estimations for $R=100$ replicated and independent datasets. Then, for each parameter we compute the mean squared error of point prediction, averaging over the $R$ replicates, the average standard errors $\mbox{se}(\hat{\rho}_{r,s})$ and the empirical coverage of the procedure counting how many times over $R$ the true $\rho_{r,s}$ is inside the obtained confidence interval. The confidence level is fixed to $\alpha =0.05$.

 In evaluating the performance of the proposed solution in terms of confidence interval estimation, it must be taken into account that we are considering the empirical estimates of the sensitivity and variability matrices provided in equations \eqref{Jhat}--\eqref{Hhat}. However, such empirical estimates may be not very accurate as discussed by \citet{cattelan}. A workaround solution proposed by the latter authors consists in estimating the sensitivity and variability matrices via Monte Carlo simulation. This solution, despite providing more reliable estimates of the Godambe's matrix, comes at an increasing computational cost and its implementation must be considered on a case-by-case basis.

To assess the inferential performance in different scenarios, we simulate data sets of different dimension $q$. Specifically we simulated data with $q=10,15$. As discussed, the number of parameter is $O(q^2)$ and hence in order to have reasonable estimates of the parameter, we need more data points for increasing dimension $q$. The sample sizes are $n=300,400,500$ and $n=400,600,800$ for $q=10$ and for $q=15$, respectively. The latent polychoric correlation matrices are correlation matrices randomly generated assuming a sparse structure with 30\% of zeroes. We fix $K=4$ and define two set of  thresholds, $(0, 0.5 , 1)^T$ and $(-1,0,1)^T$ randomly assigned to each marginal variable.

\begin{figure}
	\centering
	\subfigure[]{\includegraphics[width=0.7\textwidth]{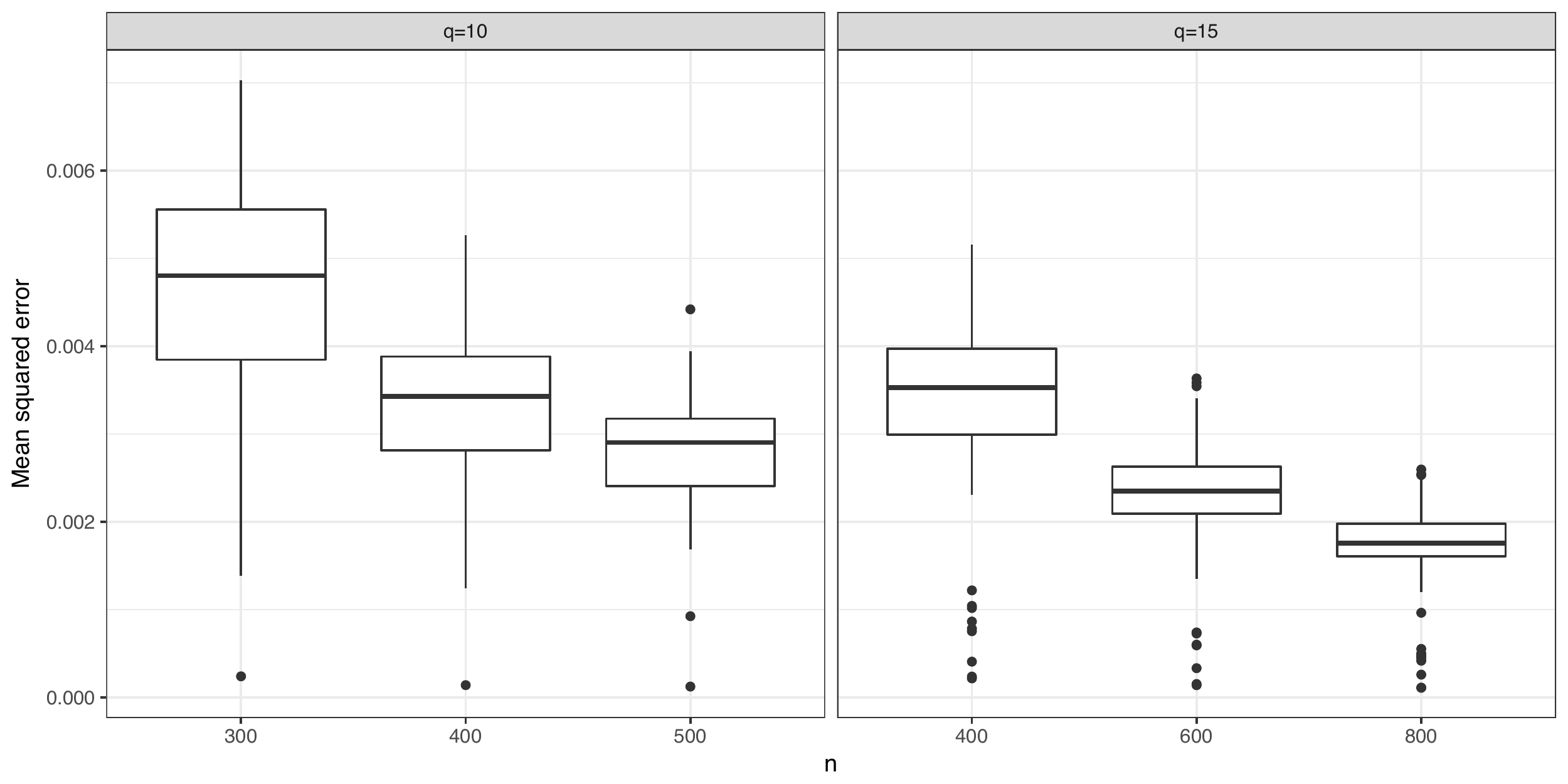}}
		\subfigure[]{\includegraphics[width=0.7\textwidth]{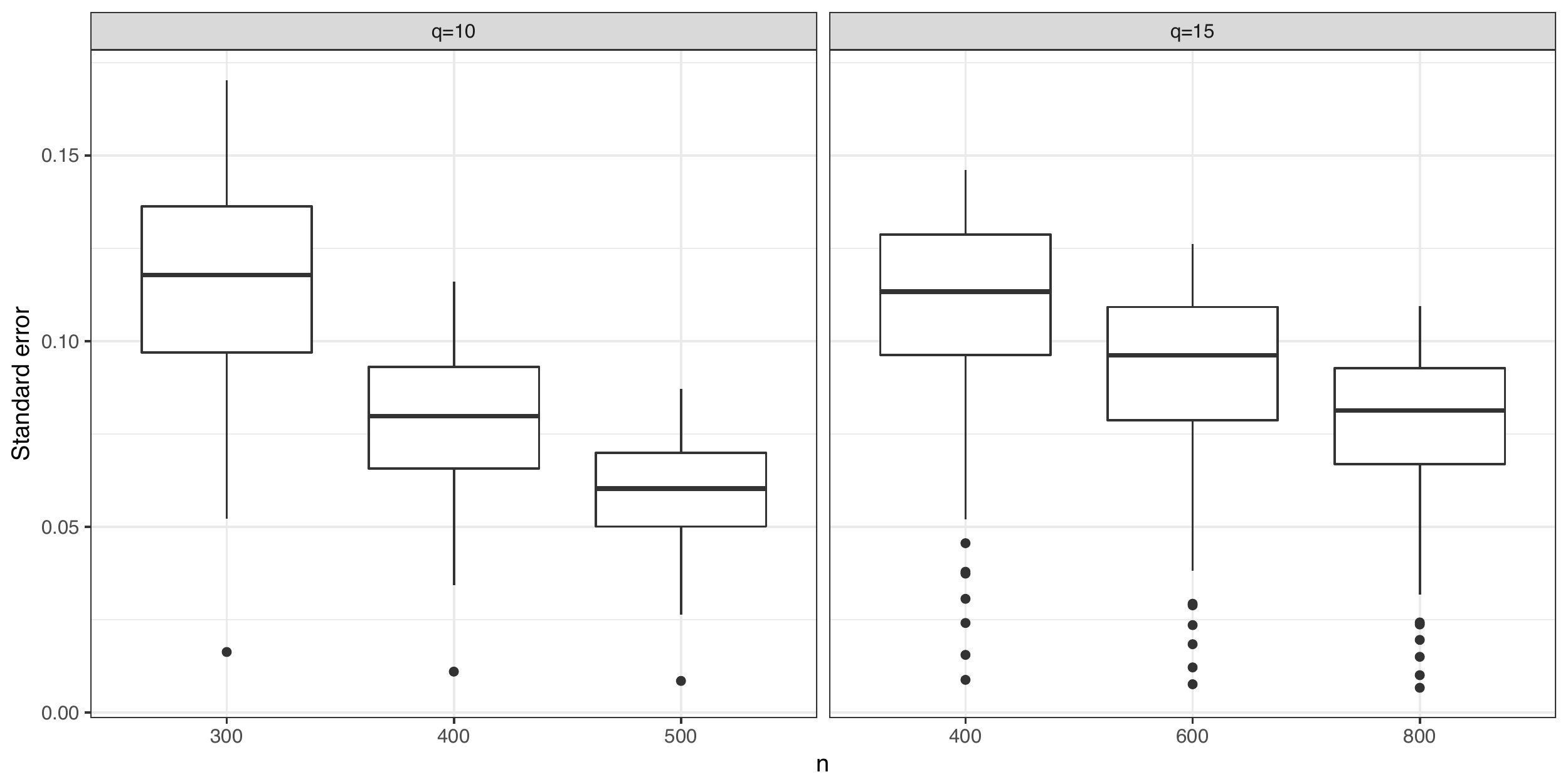}}
			\subfigure[]{\includegraphics[width=0.7	\textwidth]{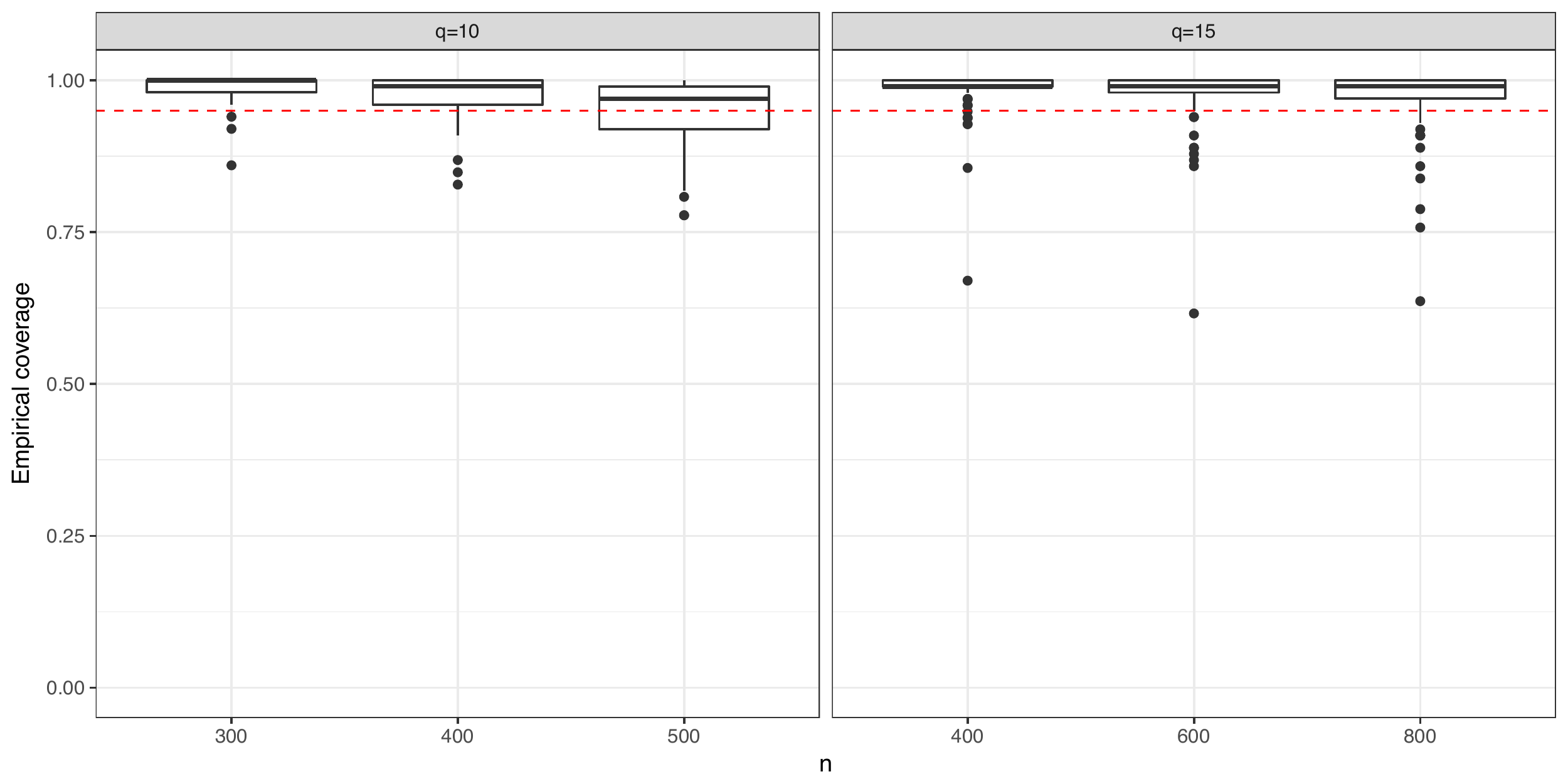}}
	\caption{Mean squared error (a), mean standard error of the estimator (b), and empirical coverage in the simulation study. The nominal confidence level ({\color{red} - - - }) in panels (c) is  0.95.}
	\label{fig:simresult}
\end{figure}

The results of the simulation study are reported in Figure~\ref{fig:simresult}. As expected, and consistently with the findings of  \cite{tonyclo} and \cite{viennesi} the mean squared errors---reported in panels (a)---are low in absolute value and substantially decrease for increasing sample size in both scenarios. The average magnitude of the standard errors also decreases for increasing sample size as can be noticed in panels (b). This is also an expected behaviour as the precision of the estimators is expected to grow with the sample size. The empirical coverage, depicted in the last panels of Figure~\ref{fig:simresult} is close to its nominal values but shows, in all cases, 
a moderate over confidence. This is probably due to the already discussed inaccuracy of the sample estimates of the variability and sensitivity matrices which may lead to larger standard errors and hence wider intervals. Despite being conservative and showing a considerable variability for each parameter, globally the empirical coverage converges to its nominal value as the sample size increases, i.e. when the averages in \eqref{Jhat}--\eqref{Hhat} are calculated with more data points.

\section{Discussion}
\label{conclusion}
In this article, we studied a pairwise likelihood approach for inference in the multivariate ordered probit model. Specifically, we derived and described the analytical expression of the pairwise score vector.  This result is of paramount interest both for point and interval estimation. Indeed the analytical expression led to a dramatic reduction of the computational costs related to the numerical maximization of the pairwise log-likelihood function via standard gradient-based numerical optimization as discussed in Section~\ref{sec:time}. 

In addition, the pairwise score allowed us to compute an empirical estimate of the Godambe matrix that we used for standard errors quantification and confidence interval estimations. Despite showing promising performance in the simulation study, the empirical estimates of the sensitivity and variability matrices provided in equations \eqref{Jhat}--\eqref{Hhat} must be used with some care. Indeed, for small sample sizes, they may be very inaccurate and a Monte Carlo estimator may be preferable \cite{cattelan}. This solution comes at an increasing computational cost which further motivates the need of the fast and reliable routine to compute the pairwise score vector based on our analytical results. 

In the specific class of models that we considered here, all the $q$ marginals have the same number of categories $K$. However, extensions to the cases in which each marginal can have different levels of categories $K_j$ are straightforward. We also assumed that the latent variables are iid but in many situations it is reasonable to assume some sort of dependence, e.g. from a set of known covariates \citep{tonyclo,viennesi}. This aspect is of dramatic interest in many application and is subject to ongoing research. 

\section*{Acknowledgement} 

%The authors are grateful for useful comments from the Managing Editor and three referees which improve the presentation of the manuscript.  
This work was initially developed by Martina Bravo as a MSc thesis in {\textsl{Stochastics and Data Science}} at the University of Torino, Italy.

%\bibliography{wileyNJD-APA}%

\bibliographystyle{elsarticle-harv} 
\bibliography{myref}

\end{document}